\RequirePackage{ifpdf}
\ifpdf 
\documentclass[pdftex]{sigma}
\else
\documentclass{sigma}
\fi

\usepackage{amsopn}

\def\ds{\stackrel{\star}{,}}
\def\dtr{\stackrel{\star}{\triangleright}}

\def\p{\partial}

\def\lb{\lbrack}
\def\rb{\rbrack}

\begin{document}

\allowdisplaybreaks

\renewcommand{\PaperNumber}{089}

\FirstPageHeading

\renewcommand{\thefootnote}{$\star$}

\ShortArticleName{Einstein--Riemann Gravity on Deformed Spaces}

\ArticleName{Einstein--Riemann Gravity on Deformed
Spaces\footnote{This paper is a contribution to the Proceedings of
the O'Raifeartaigh Symposium on Non-Perturbative and Symmetry
Methods in Field Theory
 (June 22--24, 2006, Budapest, Hungary).
The full collection is available at
\href{http://www.emis.de/journals/SIGMA/LOR2006.html}{http://www.emis.de/journals/SIGMA/LOR2006.html}}}

\Author{Julius WESS~$^{\dag^1\dag^2\dag^3}$}
\AuthorNameForHeading{J. Wess}

\Address{$^{\dag^1}$~Arnold Sommerfeld Center for Theoretical
Physics Universit\"at M\"unchen,\\
$\phantom{^{\dag^1}}$~Theresienstr.\ 37, 80333 M\"unchen, Germany}
\EmailDD{\href{mailto:wess@theorie.physik.uni-muenchen.de}{wess@theorie.physik.uni-muenchen.de}}

\Address{$^{\dag^2}$~Max-Planck-Institut f\"ur Physik, F\"ohringer
Ring 6, 80805 M\"unchen, Germany}

\Address{$^{\dag^3}$~Universit\"at Hamburg, II Institut f\"ur Theoretische Physik and DESY,\\
$\phantom{^{\dag^3}}$~Luruper Chaussee 149, 22761 Hamburg, Germany}

\ArticleDates{Received October 27, 2006, in f\/inal form November
28, 2006; Published online December 11, 2006}

\Abstract{A dif\/ferential calculus, dif\/ferential geometry and
the E-R Gravity theory are studied on noncommutative spaces.
Noncommutativity is formulated in the star product formalism. The
basis for the gravity theory is the inf\/initesimal algebra of
dif\/feomorphisms. Considering the corresponding Hopf algebra we
f\/ind that the deformed gravity is based on a deformation of the
Hopf algebra.}

\Keywords{noncommutative spaces; deformed gravity}

\Classification{83C65; 81T75; 58B34}

\begin{quote}
\it This article is based on common work with Paolo Aschieri,
Christian Blohmann, Marija Dimitrijevi\' c, Frank Meyer and Peter
Schupp {\rm \cite{grav1,twist2}}.
\end{quote}

\section{Introduction}

Gravity theories and dif\/ferential geometry have been developed
on dif\/ferential manifolds where the functions form an algebra by
pointwise multiplication:
\begin{gather}
\label{in1} \mu\{f\otimes g\} = f\cdot g .
\end{gather}
In this lecture I want to show that these theories can be
generalized by deforming this product. There are many deformations
of the pointwise product to a star product
\cite{flato,flato-a,flato-b,flato-c}; the simplest and most
discussed  is the Moyal--Weyl product \cite{weyl,weyl-a}:
\begin{gather}
\label{in2} \mu_\star\{f\otimes g\} \equiv f\star g =
\mu\big\{e^{\frac{ih}{2}\theta^{\rho\sigma} \p_\rho
\otimes\p_\sigma} f\otimes g\big\} ,
\end{gather}
where $\theta^{\rho\sigma}$ is constant antisymmetric matrix. This
product can be shown to be associative but it is not commutative.
It is def\/ined for $C^{\infty}$ functions in general as a formal
power series in $h$. Evaluated on the functions $x^\mu$ and
$x^\nu$ (\ref{in1}) yields:
\begin{gather*}
x^\mu\star x^\nu - x^\nu\star x^\mu \equiv [x^\mu \ds x^\nu ] =
i\theta^{\mu\nu}.
\end{gather*}
These, mathematically, are the canonical commutation relations of
quantum mechanics but here we postulate them for the
conf\/iguration space.

A dif\/ferential calculus on noncommutative spaces has been
developed \cite{diffcal,diffcal-a,diffcal-b}. Considering
dif\/ferentiation as a map from the space of functions to the
space of functions:
\begin{equation}
\label{in4}
\partial _\rho\colon f \mapsto  \partial_\rho f ,
\end{equation}
it can be generalized to an algebra map.

Recognizing that $f\star g$  is a function, again one f\/inds the
deformed Leibniz rule:
\begin{gather*}
\p_\rho (f\star g) = (\p_\rho f)\star g + f\star (\p_\rho g) +
f(\p_\rho \star)g .
\end{gather*}
In the case of the Moyal--Weyl product the $\star$ operation is
$x$-independent and we obtain the usual Leibniz rule. To indicate
that the derivative now is a map from the deformed algebra of
functions to the deformed algebra of functions we denote it by
$\p^\star$
\begin{gather*}
{\partial}^\star_\rho f \equiv  \partial _\rho f,\qquad
\p^\star_\rho (f\star g) = (\p^\star_\rho f)\star g + f\star
(\p^\star_\rho g) .
\end{gather*}
These equations establish a well-def\/ined dif\/ferential calculus
on the deformed space of functions. They allow us to consider
$\p^\star_\rho$ as a linear operator with the properties:
\begin{gather}
\label{in7} \p^\star_\rho\p^\star_\sigma =
\p^\star_\sigma\p^\star_\rho
\end{gather}
and
\begin{gather}
\label{in8} {\partial}^\star_\rho f = (\p^\star_\rho f) + f\star
\p^\star_\rho = (\p_\rho f) + f\star \p_\rho .
\end{gather}
The following treatment of deformed dif\/ferential geometry will
be based on the equations (\ref{in2}), (\ref{in7}) and
(\ref{in8}). It is only the equation (\ref{in4}) that def\/ines
the ordinary derivative of a function that has to be used as an a
priori input. The generalization to the deformed situation is
essentially algebraic of nature.

\section[Differential operators]{Dif\/ferential operators}

We now consider the extension of the algebra of functions
(deformed or undeformed) by the algebra of derivatives. From the
Leibniz rule (\ref{in8}) follows that there is a basis where the
derivatives are all at the right hand side of the functions. An
element of the extended algebra in this basis we call a
dif\/ferential operator \cite{grav1}.

Undeformed:
\begin{gather*}
{\cal{D}}_{\{ d\} } = \sum_{r\ge 0}d_r^{\rho_1\dots
\rho_r}\p_{\rho_1}\cdots\p_{\rho_r} .
\end{gather*}

Deformed:
\begin{gather*}
{\cal{D}}^\star_{\{d\}} = \sum_{r\ge 0}d_r^{\rho_1\dots
\rho_r}\star\p^\star_{\rho_1}\cdots\p^\star_{\rho_r}.
\end{gather*}
A dif\/ferential operator is characterized by the coef\/f\/icient
functions $d_r$. This is indicated by $\{ d\}$. We shall
frequently omit this indication and write ${\cal D}$ for a
dif\/ferential operator, with coef\/f\/icient function $d_r$ and
${\cal D}'$ for $d_r'$.

Dif\/ferential operators can be multiplied using the algebraic
properties (\ref{in1}) or (\ref{in2}) in the deformed case and the
relations (\ref{in7}) and (\ref{in8}).

The product can always be expressed in terms of dif\/ferential
operators by reordering it with the help of the Leibniz rule. In
this sense the dif\/ferential operators form an algebra in both
cases, deformed and undeformed.

There is a map of the operators ${\cal D}$ to the operators ${\cal
D}^\star$ that is an algebra morphism
\begin{gather*}
{\cal{D}}\mapsto {\cal{D}}^\star .
\end{gather*}
To def\/ine this map we let the dif\/ferential operators act on a
function $g$:
\begin{gather*}
{\cal{D}}\triangleright g = \sum_{r\ge 0}d_r^{\rho_1\dots
\rho_r}\big(\p_{\rho_1} \cdots\p_{\rho_r} g\big)
\qquad\text{or}\qquad
{\cal{D}}^\star\dtr g = \sum_{r\ge 0}d_r^{\rho_1\dots
\rho_r}\star\big(\p^\star_{\rho_1} \cdots\p^\star_{\rho_r} g\big)
.
\end{gather*}
We now ask for an operator that it, when acting on $g$, maps $g$
to the same function that we obtain by acting with the operator $
{\cal D}$ on $g$. This $\star$-operator we call
$X^\star_{\cal{D}}$
\begin{gather}
\label{diffop6} X^\star_{\cal{D}}\dtr g = {\cal D}\triangleright
g.
\end{gather}
Because $X^\star_{\cal{D}}\dtr g$ is a function we can apply
$X^\star_{\widetilde{\cal{D}}}$ once more:
\begin{gather*}
X^\star_{\widetilde{\cal{D}}}\dtr
\Big(X^\star_{{\cal{D}}}\triangleright g\Big) = \Big(
X^\star_{\widetilde{\cal{D}}}\star X^\star_{{\cal{D}}}\Big) \dtr
g.
\end{gather*}
The left hand side can also be evaluated by using (\ref{diffop6})
consecutively:
\begin{gather*}
X^\star_{\widetilde{\cal{D}}}\dtr \Big(X^\star_{{\cal{D}}}\dtr
g\Big) = X^\star_{\widetilde{\cal{D}}}\dtr \Big({\cal
D}\triangleright g\Big) = \widetilde{\cal D}{\cal D}
\triangleright g = X^\star_{\widetilde{\cal D}{\cal D}} \dtr g.
\end{gather*}
We conclude:
\begin{gather}
\label{diffop9} X^\star_{\widetilde{\cal{D}}}\star
X^\star_{{\cal{D}}} = X^\star_{\widetilde{\cal D}{\cal D}} .
\end{gather}

Multiplying $g$ pointwise with a function $f$ forms a subalgebra
of ${\cal D}$. We shall now construct the operator $X^\star _f$
explicitly for this case starting from (\ref{in1}):
\begin{gather*}
f\cdot g = \mu\{f\otimes g\} =
\mu\{e^{\frac{ih}{2}\theta^{\rho\sigma}\p_\rho
\otimes\p_\sigma}e^{-\frac{ih}{2}\theta^{\rho\sigma}\p_\rho
\otimes\p_\sigma}f\otimes g\} =\mu_\star\{
e^{-\frac{ih}{2}\theta^{\rho\sigma}\p_\rho
\otimes\p_\sigma}f\otimes g\} .
\end{gather*}
More explicitly:
\begin{gather*}
X^\star _f = \sum_{r\ge 0}\frac{1}{r!}\biggl(-\frac{i}{2}\biggr)^r
\theta^{\rho_1\sigma_1}\cdots\theta^{\rho_r\sigma_r}(\partial_{\rho_1}
\cdots\partial_{\rho_r}f)\star \partial^\star
_{\sigma_1}\cdots\partial^\star _{\sigma_r} .
\end{gather*}
This operator has the properties:
\begin{gather*}
X_f^\star \star g = f\cdot g \qquad \text{and}\qquad
X_f^\star \star X^\star_g = X^\star_{fg} .
\end{gather*}
Deformed gauge theories \cite{mex,mex-a,defgt,defgt-a} are based on these operators.

The algebra of functions with pointwise multiplication is mapped
into an algebra of deformed dif\/ferential operators.

{\samepage The algebra of dif\/feomorphisms is generated by vector
f\/ields
\begin{gather}
\label{diffop14} \xi = \xi^\mu (x)\partial_\mu ,\nonumber
\\
\lb \xi, \eta \rb = \bigl( \xi^\mu (\p_\mu \eta^\rho)
-\eta^\mu(\p_\mu \xi^\rho) \bigr)\partial_\rho =(\xi\times
\eta)^\rho \partial_\rho =  \xi\times \eta .
\end{gather}
The commutator of two vector f\/ields is a vector f\/ield again.
This is not the case for the star commutator because the
$\star$-product of two functions does not commute. The
dif\/ferential operators~$X_\xi^\star$, however, will form an
algebra under the star commutator:
\begin{gather*}
\lb X^\star_\xi \ds X^\star_\eta \rb = X^\star_{\xi\times \eta} .
\end{gather*}
This follows from (\ref{diffop9}).

}

The operator $X^\star_\xi$ is easily constructed starting from
(\ref{diffop14})
\begin{gather*}
\xi\triangleright g = \xi^\mu\p_\mu\triangleright g =
\xi^\mu(\p_\mu g) = X^\star_{\xi^\mu}\star\p^\star_\mu\dtr g .
\end{gather*}
Thus, we f\/ind
\begin{gather*}
X^\star_\xi = X^\star_{\xi^\mu}\star\p^\star_\mu .
\end{gather*}

Again the usual algebra of dif\/feomorphisms is mapped into a
subalgebra of the algebra ${\cal D}^\star$
\begin{gather*}
\xi \mapsto  X^\star_\xi, \qquad \lb X^\star_\xi \ds X^\star_\eta
\rb = X^\star_{\xi\times \eta} .
\end{gather*}
This is the starting point for the construction of a tensor
calculus on tensor f\/ields.

\section[Tensor fields]{Tensor f\/ields}

The classical theory of gravity is based on invariance under
coordinate transformations. This leads to the concept of scalar,
vector and tensor f\/ields that transform under general coordinate
transformation as follows
\begin{gather*}
\begin{array}{ll}
{\mbox{scalar: }}&  \delta_\xi\phi(x) = -\xi\phi,
\\[1ex]
{\mbox{covariant vector: }} &  \delta_\xi V_\mu(x) = -\xi V_\mu -
(\p_\mu\xi^\rho)V_\rho,
\\[1ex]
{\mbox{contravariant vector: }} &  \delta_\xi V^\mu(x) = -\xi
V^\mu + (\p_\rho\xi^\mu)V^\rho
\end{array}
\end{gather*}
and so on.

The concept of coordinate transformations is dif\/f\/icult to
generalize to deformed spaces, but the transformation laws of
f\/ields are representations of the inf\/initesimal
dif\/feomorphism algebra that we know how to deform. Thus, we will
postulate the following transformation laws for the deformed
algebra of dif\/feomorphisms
\begin{gather*}
\delta^\star_\xi\phi(x) = -X^\star_\xi \dtr\phi,
\\
\delta^\star_\xi V_\mu(x) = -X^\star_\xi\dtr V_\mu-
X^\star_{(\p_\mu\xi^\rho)}\dtr V_\rho,
\\
\delta^\star_\xi V^\mu(x) = -X^\star_\xi \dtr V^\mu+
X^\star_{(\p_\rho\xi^\mu)} \dtr V^\rho
\end{gather*}
and so on.

To construct Lagrangians we have to know how the $\star$-product
of f\/ields transforms. These products should transform as tensor
f\/ields again. E.g., the $\star$-product of two scalar f\/ields
should transform as a scalar f\/ield again
\begin{gather}
\label{tf3} \delta^\star_\xi(\phi\star\psi) = -X^\star_\xi\dtr
(\phi\star\psi).
\end{gather}
A lengthy but direct calculation shows that this is identical to:
\begin{gather}
\label{tf4} \delta^\star_\xi(\phi\star\psi) = -\mu_\star\{{\cal
F}\Delta(\xi){\cal F}^{-1} \phi\otimes\psi \}  ,
\end{gather}
where $\Delta(\xi)$ is the usual comultiplication
\begin{gather*}
\Delta(\xi) = \xi\otimes 1 + 1\otimes \xi
\end{gather*}
and therefore an element of the $\otimes$ tensor product. ${\cal
F}$ is called a twist and it is an element of the tensor product
\begin{gather}
\label{tf6} {\cal F} = e^{-\frac{ih}{2}\theta^{\rho\sigma}\p_\rho
\otimes\p_\sigma} .
\end{gather}
The right hand side of (\ref{tf3}) and (\ref{tf4}) can be
calculated in a power series expansion in $h$ and will be found to
be the same.

The advantage of the expression (\ref{tf4}) is that it links to
the formalism of deforming Hopf algebras by twists
\cite{twist1,twist1-a,twist1-b,twist1-d}. Many results
are known there. We f\/irst have to establish that ${\cal F}$
really satisf\/ies the conditions for a twist \cite{twist2}. For
(\ref{tf6}) it is the case. Then we can use the twist to deform
the Leibniz rule for arbitrary tensor f\/ields. The procedure is
as follows:

First consider the coproduct for the undeformed transformations
\begin{gather*}
\Delta(\delta_\xi) = \delta_\xi\otimes 1 + 1\otimes \delta_\xi ,
\end{gather*}
where the variations $\delta_\xi$ are expressed by dif\/ferential
operators such that
\begin{gather*}
\delta_\xi (\phi\otimes\psi) = (\delta_\xi \phi)\otimes \psi +
\phi\otimes (\delta_\xi\psi)
\end{gather*}
for any two tensor f\/ields $\phi$ and $\psi$. This coproduct can
be twisted
\begin{gather*}
\Delta_{\cal F}(\delta_\xi) = {\cal F}\Delta(\delta_\xi){\cal
F}^{-1} .
\end{gather*}
Finally the Leibniz rule becomes
\begin{gather}
\label{tf10} \delta^\star_\xi(\phi\star\psi) =
\mu_\star\{\Delta_{\cal F}(\delta_\xi) \phi\otimes\psi \} .
\end{gather}
This is not limited to scalar f\/ields but covers the
$\star$-product of all tensor f\/ields.

Finally we can convince ourselves that this Leibniz rule has the
properties demanded at the beginning of this chapter, i.e.\ that
$\star$-products of tensor f\/ields transform as tensor f\/ields.

Equation (\ref{tf10}) allows us now to consider $\star$-products
of f\/ields and to determine their transformation properties.

As a Hopf algebra the algebra of inf\/initesimal dif\/feomorphisms
is deformed!

\section[Einstein-Hilbert gravity]{Einstein--Hilbert gravity}

The Einstein--Hilbert theory of gravity can now be constructed
following its presentation in a text book.

\subsection{Covariant derivatives}

The covariant derivative of a tensor f\/ield should again
transform as a tensor f\/ield. This can be done with the help of a
connection $\Gamma$. For a covariant vector f\/ield:
\begin{gather*}
D_{\mu}\dtr V_{\nu} = \partial^\star_{\mu} \dtr V_{\nu} -
\Gamma_{\mu\nu}^{\alpha}\star V_{\alpha}.
\end{gather*}
To be a covariant derivative the connection has to transform as
follows:
\begin{equation}
\label{ehg2} \delta^\star_{\xi}\Gamma_{\mu\nu}^{\alpha} =
-X^\star_\xi \dtr \Gamma_{\mu\nu}^{\alpha} - X^\star_{(\p_\mu
\xi^\rho)}\dtr \Gamma_{\rho\nu}^{\alpha} - X^\star_{(\p_\nu
\xi^\rho)}\dtr \Gamma_{\mu\rho}^{\alpha} + X^\star_{(\p_\rho
\xi^\alpha)}\dtr \Gamma_{\mu\nu}^{\rho}
-\partial_{\mu}\partial_{\nu}\xi^{\alpha} .
\end{equation}
This can easily be generalized to arbitrary tensor f\/ields.

\subsection{Curvature and torsion}

The curvature and torsion tensors can be def\/ined as usual
\begin{gather*}
[D_{\mu} \ds D_{\nu}]\star V_{\rho} =
R_{\mu\nu\rho}{}^{\sigma}\star V_{\sigma} +
T_{\mu\nu}{}^{\alpha}\star D_{\alpha}V_{\rho} .
\end{gather*}
They can be expressed in terms of the connection:
\begin{gather}
\label{ehg4} R_{\mu\nu\rho}{}^{\sigma} =
\partial^\star_{\nu}\dtr\Gamma_{\mu\rho}^{\sigma}
-\partial^\star_{\mu}\dtr\Gamma_{\nu\rho}^{\sigma}
+\Gamma_{\nu\rho}^{\beta}\star\Gamma_{\mu\beta}^{\sigma}
-\Gamma_{\mu\rho}^{\beta}\star\Gamma_{\nu\beta}^{\sigma},
\\
T_{\mu\nu}{}^{\alpha} =
\Gamma_{\nu\mu}^{\alpha}-\Gamma_{\mu\nu}^{\alpha}.\nonumber
\end{gather}
From the transformation law of the connection (\ref{ehg2}) follows
that curvature and tension transform like tensors if the deformed
Leibniz rule (\ref{tf10}) is used.

\subsection{Metric tensor}

The relevant dynamical variable in gravity is the metric tensor.
It is introduced as a covariant symmetric tensor of rank two:
\begin{gather}
\label{ehg6} \delta^\star_\xi G_{\mu\nu} = -X^\star _\xi \dtr
G_{\mu\nu} - X^\star _{(\p_\mu\xi^\rho)}\dtr G_{\rho\nu} - X^\star
_{(\p_\nu\xi^\rho)} \dtr G_{\mu\rho} .
\end{gather}
For $\theta=0$ we identify it with the usual metric f\/ield:
\begin{gather}
\label{ehg7} G_{\mu\nu}\Big| _{\theta=0} = g_{\mu\nu} .
\end{gather}
We will see that $g_{\mu\nu}$ is the only dynamical variable in
the Einstein--Hilbert theory.

Next we have to construct the $\star$-inverse of the metric:
\begin{gather*}
G_{\mu\nu}\star G^{\nu\rho\star}  = \delta_\mu^\rho .
\end{gather*}
Let us f\/irst construct the $\star$-inverse of a function that is
invertible in the undeformed algebra:
\begin{gather*}
f\cdot f^{-1} =1 .
\end{gather*}
The star inverse $f^{-1\star}$ is def\/ined by
\begin{gather*}
f\star f^{-1\star} = 1 .
\end{gather*}
It exists as a geometric series because $f^{-1}$ exists. The
additional terms are a power series in $\theta$. To f\/ind
$f^{-1\star}$ in a compact version we start from
\begin{gather}
f\star f^{-1} =1 + {\cal O}(\theta),\nonumber
\\
\label{ehg11} \bigl( f\star f^{-1}\bigr) ^{-1\star} = \bigl( 1+
f\star f^{-1} -1\bigr) ^{-1\star} = \sum_{n=0}^\infty \bigl( 1 -
f\star f^{-1} \bigr) ^{n\star} .
\end{gather}
The star at the $n$th power means that  all the products are star
products. By def\/inition we know that
\begin{gather}
\label{ehg12} (f\star f^{-1})\star (f\star f^{-1})^{-1\star} =1 .
\end{gather}
The star multiplication is associative. We use this for equation
(\ref{ehg12}) and write it in the form
\begin{gather*}
f\star \bigl( f^{-1}\star (f\star f^{-1})^{-1\star}\bigr) = 1 .
\end{gather*}
It follows that
\begin{gather}
\label{ehg14} f^{-1\star} = f^{-1}\star (f\star f^{-1})^{-1\star}
.
\end{gather}
The factor $(f\star f^{-1})^{-1\star}$ has been calculated in
(\ref{ehg11}) as a power series expansion in $f$ and $f^{-1}$. We
insert this into (\ref{ehg14}) and f\/ind that $f^{-1\star}$ can
be expressed in $f$ and $f^{-1}$.

To invert the metric tensor we follow the analogous procedure.
\begin{gather*}
G_{\mu\nu}\cdot G^{\nu\rho} = \delta_\mu^\rho , \qquad
G_{\mu\nu}\star G^{\nu\rho\star}  = \delta_\mu^\rho
\end{gather*}
are the def\/ining equations for $G^{\mu\nu}$ and
$G^{\mu\nu\star}$. For $G^{\mu\nu\star}$ we f\/ind
\begin{gather}
\label{ehg16} G^{\mu\nu\star} = G^{\mu\rho}\star {(G\star
G^{-1})^{-1\star}}_\rho^{\ \nu} ,
\end{gather}
where $G$ and $G^{-1}$ are short for the matrices $G_{\mu\nu}$ and
$G^{\mu\nu}$ respectively. We also can show that
\begin{gather*}
(G\star G^{-1})^{-1\star} =\sum_{n \ge 0} \big(1 - {G\star
G^{-1}}\big)^{n\star} .
\end{gather*}
Because the $\star$-product is not commutative $G^{\mu\nu\star}$
will be not symmetric in $\mu$ and $\nu$. Note that we always work
with the formal power series expansions (in the sense that they do
not necessarily converge), so the solutions for the inverses
(\ref{ehg14}) and (\ref{ehg16}) are formal.

It can now be shown explicitly from the transformation law
(\ref{ehg6}) for $G_{\mu\nu}$ that $G^{\mu\nu\star}$ transforms as
a contravariant tensor of rank 2. Deformed Leibniz rule!

In formulating the ER theory we meet the determinant and the
square root of the determinant. As it is more dif\/f\/icult to
generalize the square root to a $\star$-square root we f\/irst
introduce the vielbein as the ``square'' root of the metric
tensor. It consists of four covariant vector f\/ields $E_\mu^{\
a}$ that form the metric:
\begin{gather*}
G_{\mu\nu} = \tfrac{1}{2} \bigl( E_\mu^{\ a}\star E_\nu^{\ b} +
E_\nu^{\ a}\star E_\mu^{\ b}\bigr)\eta_{ab} .
\end{gather*}
As the $\star$-product is noncommutative we have symmetrized
$G_{\mu\nu}$ explicitly. For the vielbein f\/ields we demand in
analogy with (\ref{ehg7})
\begin{gather*}
E_\mu^{\ a}\Big| _{\theta=0} = e_\mu^{\ a} .
\end{gather*}
From the  vector-like transformation properties of $E_\mu^{\ a}$
follows that $G_{\mu\nu}$ transforms like a tensor. The
determinant of the vielbein is def\/ined as follows:
\begin{gather}
\label{ehg20} E^\star = {\mbox {det}}_\star E_\mu^{\ a} =
\tfrac{1}{4!}\varepsilon^{\mu_1\dots\mu_4}\varepsilon_{a_1\dots
a_4} E_{\mu_1}^{\ a_1}\star \dots\star E_{\mu_4}^{\ a_4} .
\end{gather}
The star on $E^\star$ and ${\mbox {det}}_\star$ indicates that all
the multiplications are $\star$-multiplications. This
normalization was chosen such that
\begin{gather*}
{\mbox {det}}_\star E_\mu^{\ a}\Big| _{\theta=0} = {\mbox {det}}
e_\mu^{\ a} .
\end{gather*}
The second $\varepsilon$-tensor is necessary because the
$\star$-product is noncommutative.

The important property of the determinant is that it transforms as
a scalar density:
\begin{gather}
\label{ehg22} \delta^\star_\xi E^\star = -X^\star_\xi\dtr E^\star
- X^\star_{(\p_\mu\xi^\mu)}\dtr E^\star .
\end{gather}
This is a consequence of the transformation law of the vielbein.
This justif\/ies the def\/inition (\ref{ehg20}) of the
determinant.

We now have all the ingredients we need to proceed for the
formulation of the Einstein--Hilbert dynamics.

\subsection[Christoffel symbol]{Christof\/fel symbol}

We demand that the covariant derivative of $G_{\mu\nu}$ vanishes:
\begin{gather*}
D_\alpha G_{\beta\gamma} = \p^\star_\alpha\dtr G_{\beta\gamma} -
\Gamma_{\alpha\beta}^\rho \star G_{\rho\gamma}-
\Gamma_{\alpha\gamma}^\rho \star G_{\beta\rho} =0 .
\end{gather*}
We permute the indices, assume that $G_{\alpha\beta}$ is symmetric
and obtain by following the analogous procedure in the classical
case:
\begin{gather*}
\Gamma^{\sigma}_{\alpha\beta} = \tfrac{1}{2} \bigl(
\p^\star_\alpha\dtr G_{\beta\gamma} + \p^\star_\beta\dtr
G_{\alpha\gamma} - \p^\star_\gamma\dtr G_{\alpha\beta}\bigr)\star
G^{\gamma\sigma\star} .
\end{gather*}
The connection is entirely expressed in terms of
$G_{\alpha\beta}$. In this case we call the connection
Christof\/fel symbol.

Again the transformation law of the connection (\ref{ehg2})
follows from the transformation law of~$G_{\alpha\beta}$.

\subsection{Ricci tensor and curvature scalar}

We obtain the Ricci tensor by contracting the upper index with one
of the three lower indices of the curvature tensor (\ref{ehg4}).
As the curvature tensor is antisymmetric in the f\/irst two
indices we have only two choices left. Contracting of the second
index is a deformation of the classical Ricci tensor:
\begin{gather}
\label{ehg25} R_{\mu\nu} = R_{\mu\sigma\nu}{}^{\sigma} .
\end{gather}
The contraction $R_{\mu\nu\sigma}{}^{\sigma}$ vanishes in the
classical limit $\theta\to 0$. Nevertheless, we could add such
a~term to the Ricci tensor (\ref{ehg25}) and obtain a deformation
of the classical Ricci tensor.

We see that the deformation of the classical theory is not unique.
Terms that are covariant and vanish for $\theta\to 0$ are quite
possible. To really make the deformation unique an additional
requirement has to be added. We take simplicity and def\/ine the
Ricci tensor by (\ref{ehg25}).

The curvature scalar we def\/ine by contraction with
$G^{\mu\nu\star}$
\begin{gather*}
R = G^{\mu\nu\star}\star R_{\mu\nu} .
\end{gather*}
Again, as $G^{\mu\nu\star}$ is not symmetric this is a choice.

We can now show by starting from the tensor $G_{\mu\nu}$ that the
curvature scalar transforms as a~scalar f\/ield
\begin{gather}
\label{ehg27} \delta^\star_\xi R = -X^\star_\xi \dtr R = -\xi^\mu
(\p_\mu R) .
\end{gather}

\subsection{Lagrangian}

The curvature scalar multiplied by the determinant $E^\star$
transforms like a scalar density. From~(\ref{ehg27}) and
(\ref{ehg22}) follows:
\begin{gather*}
\delta^\star_\xi (E^\star\star R) = -\p_\mu^\star\dtr
\big(X^\star_{\xi^\mu}\dtr(E^\star\star R) \big) .
\end{gather*}

To def\/ine an action and the variational principle to f\/ind the
f\/ield equation we have to give a~def\/inition for the integral.
A possible def\/inition is
\begin{gather}
\label{ehg29} \int f = \int \textrm{d}^4 x\hspace*{1mm}  f .
\end{gather}
This integral has the trace property
\begin{gather*}
\int \textrm{d}^4 x\hspace*{1mm}  f\star g = \int \textrm{d}^4
x\hspace*{1mm}  g\star f .
\end{gather*}

A suitable action for a gravity theory on deformed spaces is:
\begin{gather*}
S_{\mbox{\tiny EH}} = \frac{1}{2}\int{\mbox{d}}^4 x\hspace{1mm}
\bigl( E^\star\star R + {\mbox{ c.c.}}\bigr) .
\end{gather*}

A reminder: By all the transformation laws of products of f\/ields
the deformed Leibniz rule~(\ref{tf10}) has to be used.

\subsection{Field equations}

The trace property of the integral (\ref{ehg29}) allows us to
def\/ine a variational principle. Vary one f\/ield after the other
but always bring it f\/irst to the left (right) hand side of the
integral by using the trace property.

\LastPageEnding

\end{document}